# Anisotropic Superconducting Gap and Elongated Vortices with Caroli-De Gennes-Matricon States in the New Superconductor $Ta_4Pd_3Te_{16}$

Zengyi Du, Delong Fang, Zhenyu Wang, Yufeng Li, Guan Du, Huan Yang[*], Xiyu Zhu, & Hai-Hu Wen[*]

Center for Superconducting Physics and Materials, National Laboratory of Solid State Microstructures and Department of Physics, Collaborative Innovation Center for Advanced Microstructures, Nanjing University, Nanjing 210093, China

Correspondence and requests for materials should be addressed to HY, HHW (emails: huanyang@nju.edu.cn, hhwen@nju.edu.cn).

**The superconducting state is formed by the condensation of a large number of Cooper pairs. The normal state electronic properties can give significant influence on the superconducting state. For usual type-II superconductors, the vortices are cylinder like with a round cross-section. For many two dimensional superconductors, such as Cuprates, $2H-NbSe_2$ etc., albeit the in-plane anisotropy, the vortices generally have a round shape. In this paper we report results based on the scanning tunneling microscopy/spectroscopy measurements on a newly discovered superconductor $Ta_4Pd_3Te_{16}$. The chain like conducting channels of $PdTe_2$ in $Ta_4Pd_3Te_{16}$ make a significant anisotropy of the in-plane Fermi velocity. We suggest at least one anisotropic superconducting gap with gap minima or possible node exists in this multiband system. In addition, elongated vortices are observed with an anisotropy of $\xi_{\|b}/\xi_{\perp b} \approx 2.5$. Clear Caroli-de-Gennes -Matricon states are also observed within the vortex cores. Our results will initiate the study on the elongated vortices and superconducting mechanism in the new superconductor $Ta_4Pd_3Te_{16}$.**

For a type-II superconductor, magnetic flux will penetrate into the bulk and form the quantized vortices when the external magnetic field exceeds the lower critical value $H_{c1}$. For most conventional superconductors, the vortices are cylinder-like with a round cross-section. For many layered superconductors, such as the cuprates[1-4], $2H-NbSe_2$ (ref. 5, 6), iron based superconductors[7-9] etc., although vortices with low dimensionality are expected, the vortices have also the round shape. Elongated vortices are expected in the Josephson vortex systems when the magnetic field is parallel to the *ab*-planes in 2D superconductors[10-13]. Occasionally, single elongated vortex may be observed on the basal plane when the supercurrent is confined by the $C_2$ like electronic state near twin-boundaries or defects[14]. Recently superconductivity with the $PdTe_2$ chains as the conducting channels[15] was discovered in $Ta_4Pa_3Te_{16}$ (ref. 16). Band structure DFT calculations reveal a mixture of 1D, 2D and 3D Fermi surfaces in this system[17]. A recent thermal conductivity measurement[18] indicates the sizable residual thermal conductivity coefficient $\kappa/T|_{T\rightarrow 0}$ and an $H^{1/2}$ magnetic field dependence under magnetic field, suggesting a nodal gap structure. It is thus curious to know what the superconducting gaps look like. Concerning the one dimensionality conduction in the $PdTe_2$-chain based superconductors, it is also desired to see whether the vortices are present, and in what form? If the Fermi surface structure contains a 1D component, the anisotropic in-plane Fermi velocity may lead to the elongation of vortex core structure. For these elongated vortices, how about the Abrikosov lattice? Furthermore, it would be very interesting to know whether the Caroli-de-Gennes-Matricon (CdGM) states[19, 20] are still expected within the vortex cores with a strong one dimensional component.

The $Ta_4Pd_3Te_{16}$ single crystal samples were grown using self-flux method. The needle-like single crystals have the typical dimensions of about $2\times0.2\times0.1$ mm$^3$. The superconducting transition occurs at about 4.5 K and is characterized by resistivity and DC magnetic susceptibility measurements, as shown in Figs. 1c and 1d respectively. The long-rod shape of the samples with shiny top-surfaces provides information on the anisotropic crystal structures. This material is a layered system with the stoichiometric composition for each individual layer along (-103) direction, so it is very easy to cleave and the charge-neutral plane with Te atoms on the top surface will be obtained. Figure 1a represents a topographic image of a cleaved surface in an area of $50\times50$ nm$^2$. According to the crystal structure, the surface termination has no other choices but Te atoms. A stripe like feature is observed, with a spatial distance between the unidirectional bright chains of about 2.5 nm. Because the Ta atoms are the largest ones among the three kinds of elements in the compound, and the Te atoms sit just above the Ta atoms in the same line viewed in the *b*-axis shown in Fig. 1b, this may be the origin of the bright chains. Through a closer scrutiny on the atomic structure as shown in the upper panel of Fig. 1b, we find that the period distance of the neighbored chains on top-surface is exactly equal to the distance of the $PdTe_2$ chains of $Ta_4Pd_3Te_{16}$. Along each chain, one can find that the atoms arrange themselves in an alternative way with the distance between two neighbor atoms of about 3.7 Å which is consistent with the lattice constant *b* of the material. Thus the *b-axis* of the crystal must be along the bright chains. It is interesting to note that between the neighboring bright chains, some kind of periodic bright spots with a larger scale than the atomic lattice parameter along *b-axis* was observed in Fig. 1a. These bright spots assembled in a rhombus lattice form may be induced by the surface reconstruction or the charge density wave modulations, which needs further investigation. When we try to match the atomic pattern with the surface Te atoms as shown in the bottom panel of Fig. 1b, we find that the measured atoms in some 2.5 nm period regions marked with red rectangle have a shift with a half lattice parameter *b* from the expectation of the structure, i.e., the exact lattice periodicity is 5 nm while the real one measured is 2.5 nm. In short summary, the atomic structure is consistent with the model configuration of the $Ta_4Pd_3Te_{16}$ phase, beside the lattice shift a half lattice constant along *b*-axis.

Scanning tunneling spectroscopy (STS) is a direct probe to detect the local density of states (DOS), which can provide key information on the superconducting gap symmetry. In Fig. 2b, we show the STS data measured at 0.45 K by having a line scan of the spectra along the blue arrowed line crossing several bright chains on the surface shown in Fig. 2a. Although here the top-surface is not atomically resolved, the distance between two neighboring bright chains is the same as that in Fig. 1a. The superconducting feature is very clear here, and the low-energy part of the spectrum seems to be very homogenous and is perfectly reproducible. The spectra present two symmetric superconducting coherence peaks at energies of about

±0.95 mV. In the superconducting state, the differential conductance at the Fermi energy shows at least a 90% decrease from the normal state, which may provide some hints on gap symmetry. The shape of the spectra near the Fermi energy is reminiscent of that in Cuprates[21] or Chevrel phase family of superconductor $PdMo_6S_8$ (ref. 22). Before this work, the measurements of thermal conductivity in $Ta_4Pd_3Te_{16}$ suggest that there may exist nodes in the gap functions[18]. However, theoretical calculation shows that $Ta_4Pd_3Te_{16}$ may be a conventional *s*-wave superconductor[17] with Cooper pairs arising from the p-orbital electrons. In order to classify this point, we present a typical normalized STS spectrum measured at 0.45 K divided by the one taken at 5 K and show in Fig. 2c and d (symbols). Meanwhile we fit the data with several scenarios of superconducting gaps based on the Dynes model[23], these include a single isotropic *s*-wave gap, a single *d*-wave gap, an anisotropic *s*-wave gap, as shown in Fig. 2c and d. Since there are multibands in the compound $Ta_4Pd_3Te_{16}$ (ref. 17), we also used two components ($s_1+s_2$ or $s+d$) of differential conductivity with each containing a single gap function (either *s*- or *d*-wave), instead of using one component but with a mixture of two gaps (see Supplementary Information SI-I). We must emphasis that all the fittings in this Letter are the optimized ones yielding the reliable parameters. The results based on *s*-wave and *d*-wave fitting are plotted together with the experimental data in Fig. 2c. For the *s*-wave fitting, as shown by the green solid line, the calculation fails to track the low energy line shape, which always displays a sharper drop of local DOS near the Fermi energy compared with the experimental data. On the other hand, the fitting with a single *d*-wave gap, although has a better global fit to the STS curve, but generates a "V-shape" feature near the bottom, which also deviates from the experimental data. Because the single isotropic *s*-wave and *d*-wave pairing symmetry cannot appropriately interpret our data, we use an anisotropic *s*-wave gap function to simulate the data. The material has the $PdTe_2$ chains as the conducting channels, so we use a two-fold-symmetric anisotropic *s*-wave function $\Delta(\theta) = \Delta_1 + \Delta_2 \cos 2\theta$ instead of a four-fold-symmetric gap function. The best fit to the spectrum leads to $\Delta_1$ = 0.644 meV and $\Delta_2$ = 0.276 meV. For details of the fitting one is referred to the Supplementary Information. As shown in Fig. 2d, the anisotropic *s*-wave model can fit the data quite nicely, both near the bottom and the coherence peaks. The gap function is shown in the inset of Fig. 2d in a blue dumbbell like shape with a minimum value 0.37 meV and maximum value 0.92 meV. Taking the maximum gap value we determined $2\Delta_{max}/k_BT_c \approx 4.6$. We also fit the spectra with *s+d* waves and present in Fig. S1b, the fitting is as good as the one with anisotropic *s*-wave (SI-II). So we argue that the superconducting gap is highly anisotropic, or even there exist nodes or gap zero on the superconducting gap(s).

Magnetic vortices appear for type-II superconductors when a magnetic field is applied. The vortex core size can roughly give the coherence length *ξ*. Next we focus on the measurements under an applied magnetic field. Since the upper critical field $H_{c2}$ perpendicular to the cleavage surface is about 3 T, in order to maintain a less-suppressed superfluid density outside the vortex core, we applied a magnetic field of 0.8T with orientation perpendicular to the cleavage surface. Figure 3a shows a 2D mapping of the zero bias conductance over an area of 180×180 nm². In order to visualize the vortex more clearly, we filled out the signal associating with the bright chains (see SI-II). Strikingly, one can see that the vortex is elongated along the *b*-axis, i.e., the typical size is around 45 nm in the *b-axis* and 22 nm vertical to the *b-axis*, which may reveal the anisotropy of electronic properties in the cleavage plane. The average flux per vortex calculated from our data is about $1.99 \times 10^{-15}$ Wb, being close to the single magnetic flux quanta $2.07 \times 10^{-15}$ Wb. It is known that the coherence length is proportional to Fermi velocity and inversely proportional to the gap amplitude, expressed as $\xi = \hbar v_F / \pi \Delta$. According to the theoretical calculation of the Fermi surface[17] and the gap anisotropy in this Letter, the elongation can be understood qualitatively. In addition, we find that the Abrikosov lattice is also distorted along the *b*-axis, but still with a basic triangle lattice. A close scrutiny can find that the three angles enclosed by the three neighbor vortices are: 45°, 74°, and 61°, as highlighted by red triangle in Fig.3a. At a position with more symmetric vortex structure, we find the three angles of: 48°, 66°, and 66°. An elongation of the vortex lattice along *b*-axis is evidently observed.

Figure 3b displays a series of the spectra taken along the arrowed line crossing vortex centre as shown in Fig. 3c. The apparent CdGM bound state peak is clearly observed around the vortex core centre. Away from the vortex core center, the bound state disappears and the spectrum evolves continuously towards outside the vortex. When we divide the STS measured at the vortex core centre by that away from vortex centre, the CdGM state (Fig. 3d) becomes more obvious and a peak locates around the Fermi energy. On the other hand, the superconducting spectra weight at Fermi energy decreases less than 30% and the superconducting coherence peaks are suppressed dramatically outside the vortex core, indicating that the supercurrent outside the vortex core may smear up the gapped feature through the Doppler shift effect[25, 26] if the gap has a nodal or highly anisotropic structure in $Ta_4Pd_3Te_{16}$. We will further address this issue in the following.

In order to evaluate how strong the anisotropy of the vortex is and the superfluid distribution around a vortex, we measured the spectra far away from the vortex core as shown in Fig. 4b with an external field of 0.8T, and with the magnetic field released to zero as shown in Fig. 4a. The two set of data were measured by going through exactly the same trace. It is clear that the DOS at zero bias dropped more than 80% when the field is zero, but it drops only about 10% when magnetic field is applied. This is counterintuitive for an *s*-wave superconductor, since the superconducting order parameter will be established quickly outside the vortex core with a distance of about $\xi$. However, for a superconductor with strong gap anisotropy or nodes, the Doppler shift[25,26] will induce a finite DOS at $E_F$ in the region $\lambda_L > r > \xi$ (*r* the radial distance from the core center, $\lambda_L$ is the London penetration depth). The strong suppression of the superconducting coherence peak and significant lifting of the ZBC far away from the vortex core centre certainly suggests a gap minimum or gap zero point on the gap function. To extract the superconducting coherence length $\xi$, the radial dependences of the vortex-induced ZBC measured along *b*-axis and perpendicular to the *b*-axis (inset in Fig. 4c and d) are normalized to unity at the vortex centre and plotted in Fig. 4c and Fig. 4d. Then an exponential decay law[9,24] is fitted to the data. We find an average coherence length $\xi_{//b}$ = 20.6 nm, $\xi_{\perp b}$ = 8.2 nm, and the anisotropy $\xi_{//b} / \xi_{\perp b} \approx 2.5$.

For many type-II superconductors, the in-plane electronic property may have some anisotropy, but it is quite rare to see an elongated vortex. The significant elongation of the vortex in the present sample is remarkable. According to the DFT calculation[17], the system contains several Fermi pockets or sheets with a clear one dimensional feature from β and γ sheets. Unfortunately, it has no report so far about how large the in-plane anisotropy of the Fermi velocity is. Furthermore there is no any study up to now about the magnitude of the gap value and anisotropy. Our results here will help to resolve these issues. Finally, the results may also initiate the interesting trend for studying the vortex physics, both statically and dynamically. For an elongated vortex and distorted Abrikosov lattice, the vortex pinning force and the Bardeen-Stephen dissipation[27] coefficient will be also reconsidered, the vortex moving manner along *b*-axis and perpendicular to *b*-axis will certainly be different. Our discovery about the elongated vortex and the gap anisotropy stimulate the study on the new superconductor $Ta_4Pa_3Te_{16}$ and may open a new area for the study of vortex motion and phase diagram with elongated vortices.

**Methods**
I.  **Sample growth and characterization**

Single crystals $Ta_4Pd_3Te_{16}$ were grown by a self-flux method, which starts from Ta (99.9%), Pd (99%), and Te (99.999%) mixed in the mole ratio Ta:Pd:Te=2:3:15. The synthesis is similar to a method reported previously[15]. The mixed powders were thoroughly ground and then sealed in an evacuated quartz tube. It was heated up to 950 °C in 20 hours and held at this temperature for 1 day, followed by cooling to 650 °C in 60 hours and finally cooling down to room temperature by shutting off the power of the furnace. DC magnetization measurements were carried out with a SQUID-VSM-7T (Quantum Design). The electrical resistivity was measured by the standard four-probe method with current applied along the *b*-axis with a physical property measurement system (PPMS, Quantum Design).

II. **STM measurements**

The STM/S measurements were performed with an ultra-high vacuum, low-temperature, and high-magnetic-field scanning probe microscope USM-1300 (Unisoku Co., Ltd.). The samples were cleaved at room-temperature in ultra-high vacuum with a base pressure about $1\times10^{-10}$ torr. In all STM/STS measurements, tungsten tips were used. The tips were treated by in situ e-beam sputtering and calibrated on a single crystalline Au(100) surface. To lower down the noise of the differential conductance spectra, a lock-in technique with an ac modulation of 0.1 mV at 987.5 Hz was used.

**Acknowledgements:** This work was supported by the Ministry of Science and Technology of China (973 projects: 2011CBA00102, 2012CB821403), NSF of China and PAPD.

**Author Contributions** The samples were grown by Y.F.L. and X.Y.Z. The transport measurements were done by Y.F.L., H.Y. and H-H.W. The low-temperature STM/STS spectra measurements were performed by Z.Y.D., D.F., Z.W., H.Y. & H-H.W. H-H.W. coordinated the whole work. H-H.W, Z.Y.D and H.Y. wrote the manuscript, which was supplemented by other co-authors. All authors have discussed the results and the interpretation.

**Author Information** The authors declare no competing financial interests.

**Figures and Captions**

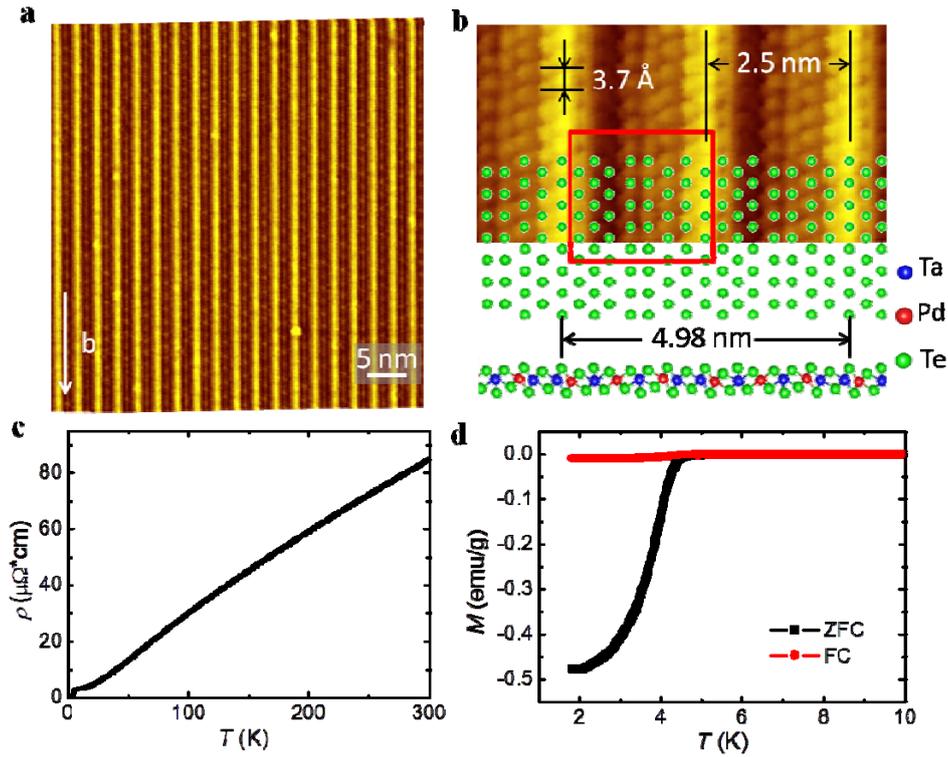

**Figure 1 | STM topography and superconducting transition of $Ta_4Pd_3Te_{16}$.** **a**, An atomically resolved topography at bias voltage $V_{bias}$ = 100 mV and tunneling current $I_t$ = 50 pA. **b**, A zoom-in view of the image in **a** and the corresponding crystal structure projected along the [-1 0 3] (upper) and [010] (lower) directions. $V_{bias}$ = 30 mV, $I_t$ = 100 pA. **c,d**, Temperature dependence of resistivity and magnetization after zero-field cooling (ZFC) and field cooling (FC) at 10 Oe.

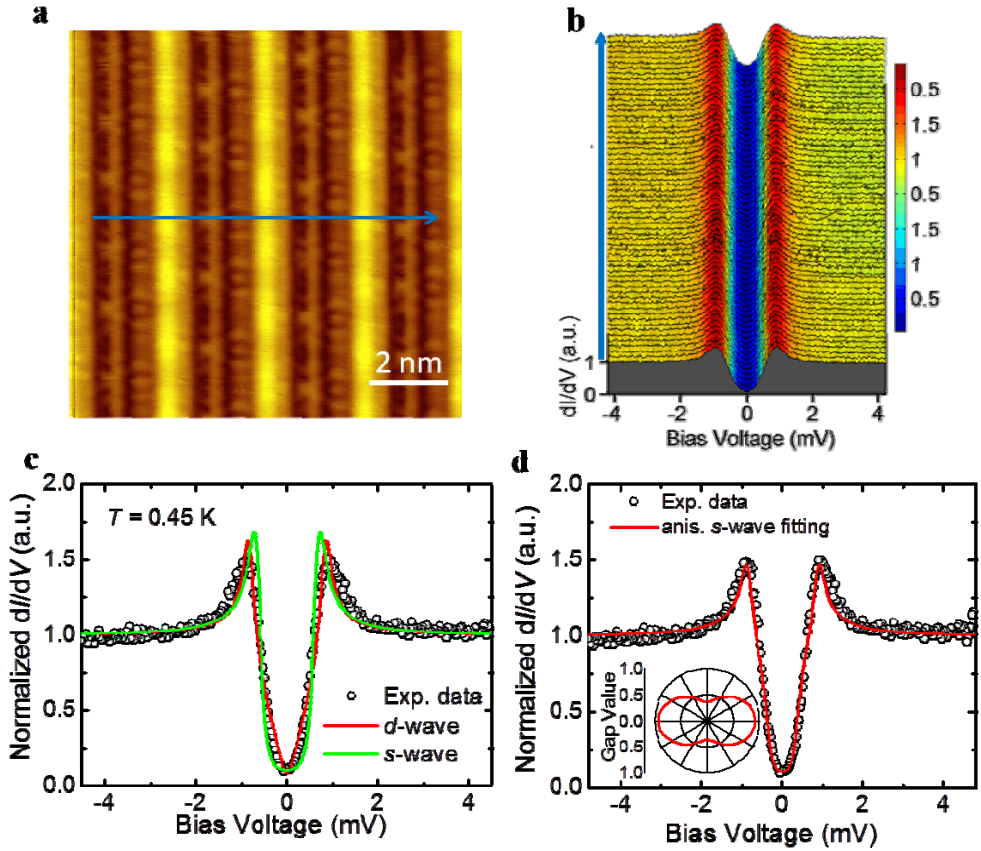

**Figure 2 | STS spectra and theoretical fitting with different gap functions. a**, Topographic image ($V_{bias}$ = 100 mV, $I_t$ = 50 pA). The blue arrow indicates the trace on which the tunneling spectra shown in **b** were measured. **b**, The spatially resolved tunneling spectra d$I$/d$V$ versus $V$ at 0.45 K. **c**,**d**, Fitting results to a typical STS spectrum at 0.45 K normalized by the one measured in the normal state (at 5 K). The symbols represent the experimental data, and the colored lines are the theoretical fits to the data with the Dynes model with *d*-wave, *s*-wave and an anisotropic *s*-wave gap, respectively. The inset in **b** shows a twofold symmetric gap function $\Delta = 0.644 + 0.276\cos 2\theta$ (meV) yielded from the fitting with an anisotropic *s*-wave gap.

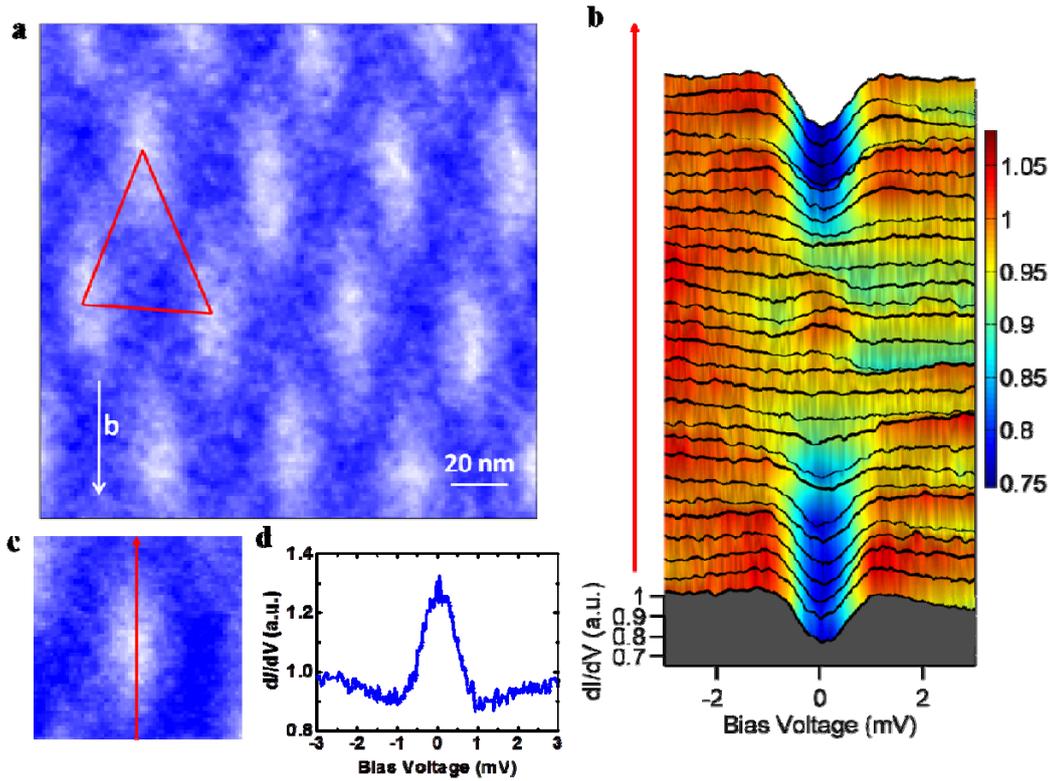

**Figure 3 | Observation of elongated vortices and the CdGM states at 0.8 T. a**, 2D mapping of differential conductance measured at 0.45 K and 0.8 T in $Ta_4Pa_3Te_{16}$. It is clear to see the vortex array composed of elongated vortices. The distorted Abrikosov hexagonal lattice elongates in *b*-axis, and a typical triangle connecting three neighbored vortices has three internal angles of 45°, 74°, and 61°. **b**, The spatially resolved tunneling spectra measured across a vortex core. The red arrow indicates the trace on which the tunneling spectra shown in **c**. **d,** A tunneling spectrum measured at the center of a vortex normalized with the one measured far away from the vortex centre.

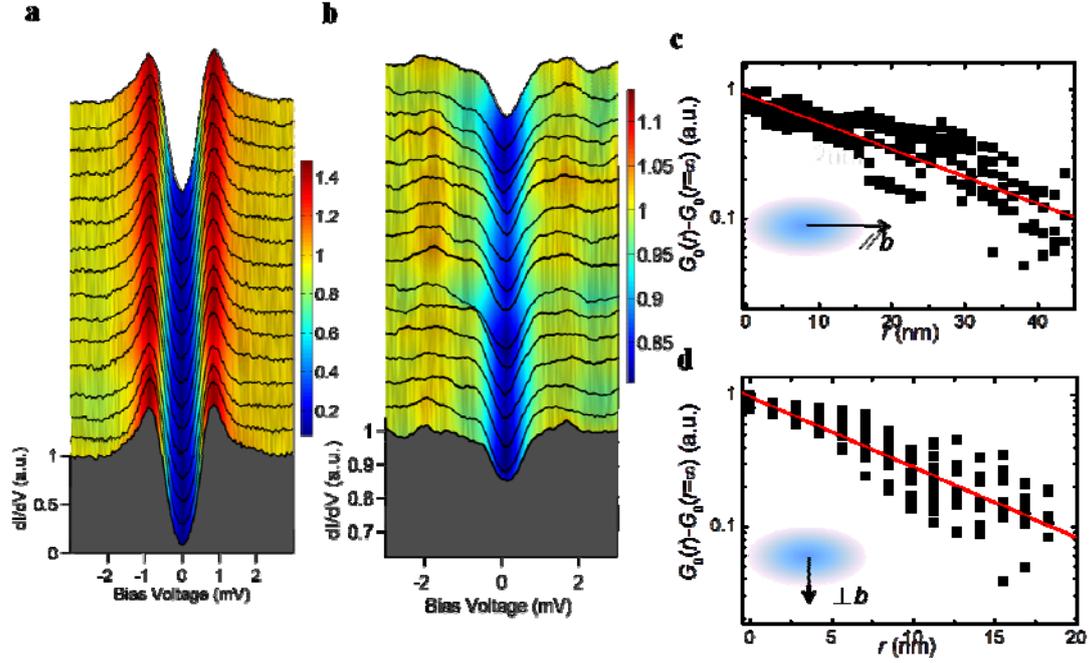

**Figure 4 | Significant lifting of the zero-bias conductance at a magnetic field and determination of the in-plane anisotropy of coherence length. a,b**, The tunneling spectra measured along the same line at zero magnetic field and 0.8 T far outside the vortex core at 0.45 K, respectively. **c,d**, Spatial dependence of the differential conductance in semi-log plots across several different vortices along the long axis and short axis. The experimental data are fitted by the exponential decay formula (red lines), which leads to an average coherence length of $\xi_{//b}$ = 20.6 nm and $\xi_{\perp b}$ = 8.2 nm at the two perpendicular directions.

# Supplementary Information

## I. STS spectra and theoretical fitting with different gap functions

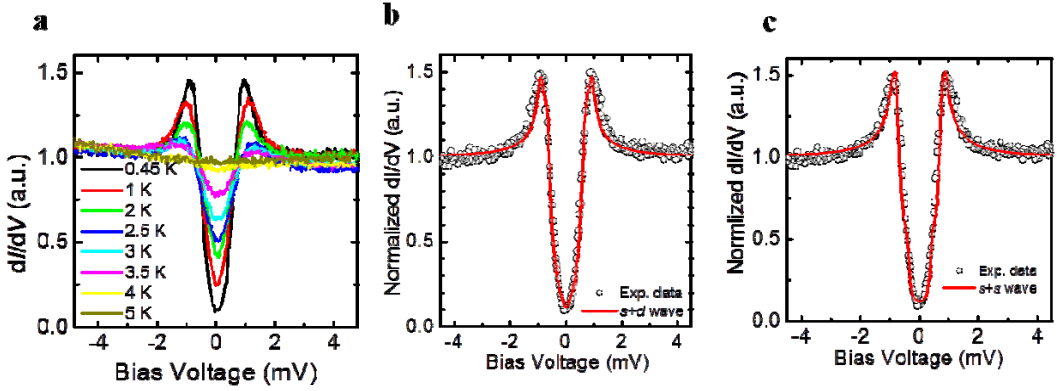

**Figure S1 | Temperature dependence of tunneling spectra and theoretical fitting. a**, Temperature dependence of tunneling spectra measured from 0.45K to 5K. **b,c** Experimental dI/dV curve and the fitting curve with different gap functions in $Ta_4Pa_3Te_{16}$ single crystal at 0.45 K. The symbols represent the experimental spectra at 0.45 K normalized by the one measured in normal state (at 5 K), the solid line in **b** shows for *d+s* wave, that in **c** shows for $s_1+s_2$ wave fitting based on the Dynes model.

Temperature dependent tunneling spectra are displayed in Fig. S1**a** from 0.45 K to 5 K (above the critical temperature of 4.5 K). In order to get a more quantitative understanding of the superconducting parameter, we use the Dynes model to fit a typical STS spectrum at 0.45 K normalized by the one measured in the normal state (at 5 K). Dynes noted that the superconducting DOS could be generalized to take into account a finite quasiparticle lifetime by writing

$$N_S(E, \Gamma) = \text{Re}\left(\frac{\varepsilon - i\Gamma}{\sqrt{(\varepsilon - i\Gamma)^2 - \Delta^2}}\right), \quad (S1)$$

So the tunneling current measured with STS can be written as

$$I(V) = \frac{1}{2\pi}\int_{-\infty}^{+\infty} d\varepsilon [f(\varepsilon) - f(\varepsilon + eV)] \cdot \text{Re}\left(\frac{\varepsilon + eV + i\Gamma}{\sqrt{(\varepsilon + eV + i\Gamma)^2 - \Delta^2}}\right). \quad (S2)$$

Here f(ε) is the Fermi function, and Γ is the inverse quasiparticle lifetime. We firstly performed fits to our experimental data using two single-band models, namely $\Delta = \Delta_0$ for an isotropic *s*-wave gap model and $\Delta(\theta) = \Delta_0|\cos 2\theta|$ for a *d*-wave gap model. The results have been shown in Fig.2c. Concerning the structure of the material, a two-fold-symmetric anisotropic s-wave gap function is also used to fit the spectra. In this case, the gap function can be written as $\Delta(\theta) = \Delta_1 + \Delta_2 \cos 2\theta$. The best fit is shown in Fig.2d, which leads to $\Delta(\theta) = 0.644 + 0.276\cos2\theta$ with the maximum and minimum gaps of 0.92 and 0.37 meV, respectively.

When using a simplified two-gap model to fit the normalized spectra, the combined deferential conductivity can be constructed as $G = pdI_1/dV + (1-p)dI_2/dV$, where $I_{1(2)}(V)$ is the tunneling current contributed by the two gap $\Delta_{1(2)}$ and $p$ is the related spectral weight of each band contributed to the tunneling current. Each $I_{1(2)}(V)$ can be described by equation S2. The fitting curve obtained by using two-band model ($s_1+s_2$ or $s+d$, each band with 50% weight) has been displayed in Fig. S2b. In the case of $d+s$, the theoretical curve can also describe the experimental data quite well.

## II. Treatment of vortex lattice with Fourier transform analysis procedure

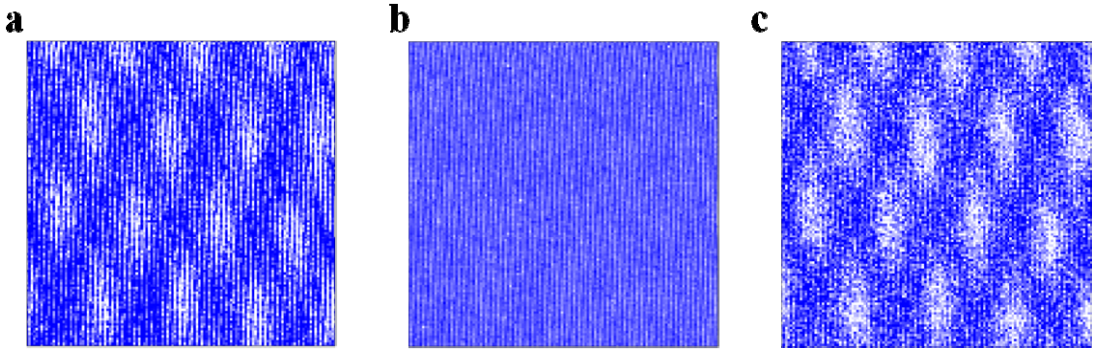

**Figure S2 | Vortex image measured at 0.45 K and 0.8 T after the treatment. a,b.** The raw data of deferential conductance image $g(\vec{r}, E)$ of $E = 0$ and $E = 0.95$ meV in an area of $180 \times 180$ nm$^2$, respectively. **c.** The difference of the deferential conductance image by taking $g(\vec{r}, E = 0 \text{ meV}) - g(\vec{r}, E = 0.95 \text{ meV})$.

We perform two steps of treatment to the raw data of vortex image shown in Fig. S2a in order to obtain a clear vortex image. Firstly, we subtracted the mapping of local conductance measured at zero voltage (Fig. S2a) by the one at 0.95 mV (Fig. S2b). After this process as shown in Fig. S2c, the vortices look clearer than in the original image, but the influence of the bright chains along $b$-axis still remains. Secondly, we deducted the peaks caused by the bright chains in the Fourier transformation pattern from Fig. S2c, then filled out the high frequency noise by performing a low-pass-filter, and finally inversed the Fourier transformation and got a clear image of vortices which are shown in Fig. 3a.